**Bandwidth Control and Symmetry Breaking in a Mott-Hubbard Correlated Metal**

*Lishai Shoham[1*], Maria Baskin[1], Tom Tiwald[2], Guy Ankonina[1], Myung-Geun Han[3], Anna Zakharova[4], Shaked Caspi[1], Shay Joseph[5], Yimei Zhu[3], Isao H. Inoue[6], Cinthia Piamonteze[4], Marcelo J. Rozenberg[7], and Lior Kornblum[1]**

1) L. Shoham, Dr. M. Baskin, Dr. G. Ankonina, S. Caspi, Prof. L. Kornblum

Andrew and Erna Viterbi Department of Electrical and Computer Engineering, Technion – Israel Institute of Technology, Haifa 3200003, Israel

E-mail: lishai@campus.technion.ac.il , liork@technion.ac.il

2) Dr. T. Tiwald

J. A. Woollam Co., Inc. 645 M Street, Suite 102, Lincoln, NE 68508, USA

3) Dr. M. G. Han, Dr. Y. Zhu

Condensed Matter Physics and Materials Science, Brookhaven National Laboratory, Upton, NY 11793, USA

4) Dr. Anna Zakharova, Dr. C. Piamonteze

Swiss Light Source, Paul Scherrer Institute, CH-5232 Villigen PSI, Switzerland

5) Dr. S. Joseph

Rafael Ltd., P.O. box 2250, Haifa 3102102, Israel

6) Dr. I. H. Inoue

National Institute of Advanced Industrial Science and Technology (AIST), Central 5, Tsukuba 305-8565, Japan

7) Prof. M. J. Rozenberg

Université Paris-Saclay, CNRS, Laboratorie de Physique des Solides, 91405 Orsay, France

In Mott materials strong electron correlation yields a spectrum of complex electronic structures. Recent synthesis advancements open realistic opportunities for harnessing Mott physics to design transformative devices. However, a major bottleneck in realizing such devices remains the lack of control over the electron correlation strength. This stems from the complexity of the electronic structure, which often veils the basic mechanisms underlying the correlation strength. Here, we present control of the correlation strength by tuning the degree of orbital overlap using picometer-scale lattice engineering. We illustrate how bandwidth control and concurrent symmetry breaking can govern the electronic structure of a correlated $SrVO_3$ model system. We show how tensile and compressive biaxial strain oppositely affect the $SrVO_3$ in-plane and out-of-plane orbital occupancy, resulting in the partial alleviation of the orbital degeneracy. We derive and explain the spectral weight redistribution under strain and illustrate how high tensile strain drives the system towards a Mott insulating state. Implementation of such concepts will drive correlated electron phenomena closer towards new solid state devices and circuits. These findings therefore pave the way for understanding and controlling electron correlation in a broad range of functional materials, driving this powerful resource for novel electronics closer towards practical realization.

# 1. Introduction

Recent advancement in the synthesis of high quality films and heterostructures opens promising avenues towards their application in new types of functional devices.[1] One of the most attractive implementations remains the Mott field effect transistor, dubbed as the MottFET,[2–4] where the vision is to use an external stimulus, such as an electric field, to drive a Mott material across a metal to insulator transition (MIT). However, these promises have yet to be realized in practical scenarios. Development of efficient tuning mechanisms for these materials will increase their technology potential and contribute to the development Mott devices such as the MottFET.

Correlated materials are a class of materials where strong electron correlations, defined as the ratio U/W between the electron Coulomb repulsion energy (U) and the width of the one-electron band (W),[5] dictates the electronic structure. Sufficiently strong electron correlation can result in a formation of the lower and upper Hubbard bands (LHB, UHB, respectively), by transferring spectral weight from the vicinity of the Fermi level. Increasing the electron correlation strength, therefore, can drive an integer-filling Mott-Hubbard material towards a Mott insulating state.[6–8] In the intermediate state, when U/W ~ 1, the LHB and UHB cancoexist with the remaining spectral weight around the Fermi level, termed the quasiparticle (QP) peak.

Bandwidth control is an attractive route for tuning the electronic structure of correlated materials by changing the hopping amplitude of the carriers, determined by the degree of orbital overlap.[5] This concept can be realized by structural distortions of the unit cell, e.g., by applying chemical[9] or hydrostatic[10] pressure, or via epitaxial strain.[11–15] For example, in the $Ca_{1-x}Sr_xVO_3$ system, the value of x controls the V-O-V bond angle without changing the vanadium valance state, which was shown to result in bandwidth modulation.[16,17] In the case of epitaxial strain, the asymmetric nature of the structural distortion typically results in orbital symmetry breaking between the in-plane and out-of-plane orbitals.

Transition metal oxides provide a rich playground for correlated electron phenomena, particularly for Mott physics. $SrVO_3$ (SVO) has been a long-time prototype for strongly correlated metals owing to its relative simplicity; this is due to its cubic perovskite structure, paramagnetism, and single valence electron in the V-3*d* band.[8,18–23] As predicted by theoretical studies[24,25] and shown by spectroscopy,[25–28] the V-3*d* band for SVO presents a three-peak structure consisting of the LHB, QP peak and UHB. Nevertheless, as a correlated metal, most of its spectral weight is located around the QP peak. Recent improvement in thin film growth methods[29–31] opens the door for new possibilities, such as extending early bandwidth control attempts to a biaxial landscape.

The effect of biaxial epitaxial strain on the electronic landscape of SVO consists of two key contributions, as predicted by theory.[7] On the one hand, structural changes in the unit cell decrease (increase) the hopping amplitude under tensile (compressive) strain. On the other hand, symmetry breaking induces a crystal field split, partially alleviating the degeneracy of the V-$3d$ $t_{2g}$ orbitals, which constitute the backbone of the $t_{2g}$ subbands of the V-$3d$ band. The degeneracy removal reduces the number of bands available for electron delocalization at the ground state,[32,33] regardless of the applied strain. Thus, under tensile strain, the concomitant contributions are expected to increase the correlation strength. Conversely, for compressive strain, these contributions counteract each other, and the effect on the correlation strength, is expected to be small. Altogether, biaxial strain is a powerful knob for controlling the electronic structure. However, the complexity of many systems often obscures the underlying mechanisms behind the correlation strength, which may explain why experimental realization of these concepts remains challenging.

In this work, we directly control the electron correlation strength of SVO by manipulation of the V-O bond length via biaxial epitaxial strain. Leaning on the theoretical foundations and predictions of Ref. [7], we start by confirming the V-O bond length modulation and strain-selective crystal field splitting (sections 2.1 and 2.2, respectively). We then apply optical conductivity analysis to illustrate how the hopping amplitude and crystal-field splitting variation reshape the electronic structure (section 2.3). We rule out any octahedral rotation using atomic resolution imaging and electron diffraction (section 2.4). The choice of the simple SVO as a model allowed us to obtain a clear picture of these concurrent effects, paving the way to generalize the results to a broad range of more complex systems.

## 2. Results and discussion

### 2.1. Bond-length modulation

Epitaxial strain, which stems from a mismatch between the film and substrate, can tune the SVO electronic structure by mechanical modulation of the V-O bond length. To understand these mechanisms, we grew epitaxial SVO films (nominally 25-30 nm thick) on the following substrates: SrTiO$_3$ with 1.7% mismatch (STO, 1.7%), LaGaO$_3$ (LGO, 1.2%), (LaAlO$_3$)$_{0.3}$(Sr$_2$TaAlO$_6$)$_{0.7}$ (LSAT, 0.7%), and LaAlO$_3$ (LAO, -1.4%). The mismatch is defined as 100%·($a_{sub}$-$a_{SVO}$)/$a_{sub}$, where $a_{sub}$ and $a_{SVO}$ are the bulk (or pseudocubic) lattice parameters of the substrate and SVO, respectively (**Figure 1a**). Bulk (or unstrained) SVO has a lattice parameter of 3.843 Å.[18], and a positive (negative) mismatch value translates to tensile (compressive) strain in this notation. The films exhibit high crystalline quality (Figure S1) and low defect densities, evident by the high residual resistivity ratios (RRR) ranging from 4.5 to 15.3 (Figure S2). Reciprocal space maps (RSMs, **Figure 1b**) confirm that all films are fully strained to their substrates. Thus, the nominally-cubic SVO unit cell undergoes tetragonal distortion under the biaxial in-plane epitaxial strain applied by the substrate.

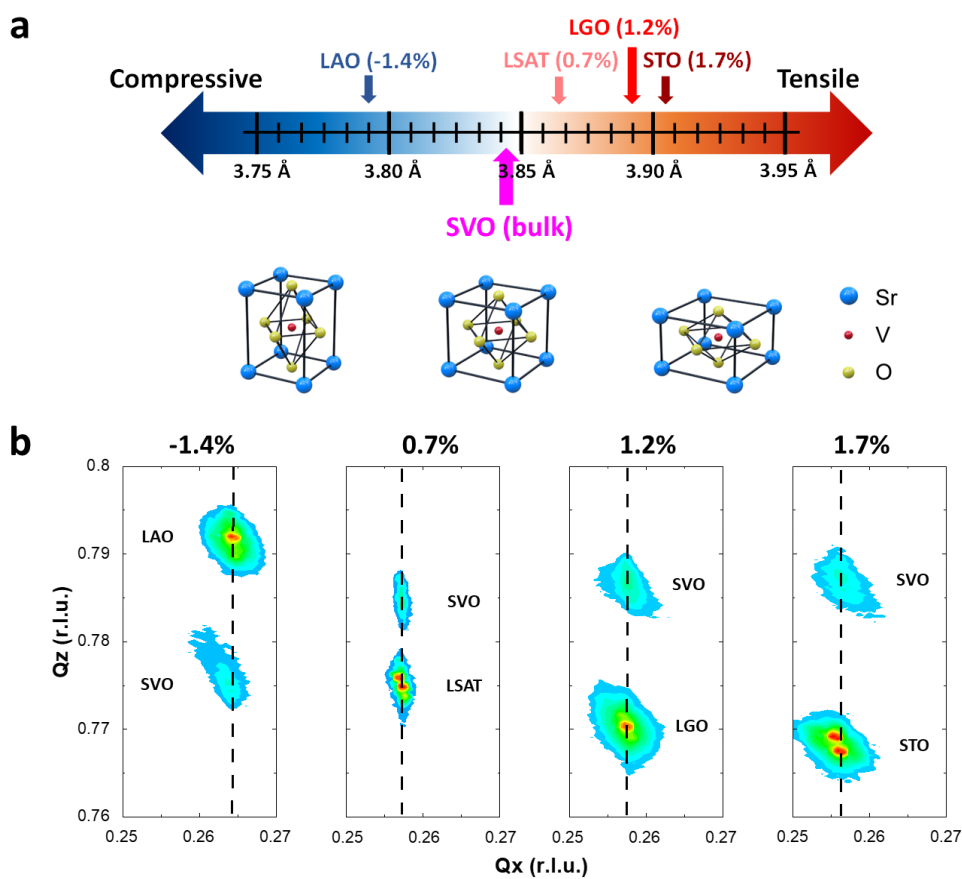

**Figure 1.** (a) Comparison of the (cubic\pseudocubic) lattice parameters of (bulk) SrVO$_3$ (SVO) and the various substrates used in this study to strain it: SrTiO$_3$ (STO), LaGaO$_3$ (LGO), (LaAlO$_3$)$_{0.3}$(Sr$_2$TaAlO$_6$)$_{0.7}$ (LSAT), and LaAlO$_3$ (LAO). (b) Reciprocal space maps (RSMs) of the strained SVO films grown on different substrates indicate coherent growth. All measurements were taken in the cubic\pseudocubic (0 1 3) orientation.

From the structural analysis, we conclude that the V-O bond length directly manifests the biaxial in-plane strain (octahedral rotations are ruled out in section 2.4). Under tensile strain, the in-plane V-O bond length increases, reducing the V 3d – O 2p orbital overlap and decreasing the hopping amplitude. The opposite occurs for compressive strain, where the strain-induced bond length shortening increases the orbital overlapping. Unlike several other $ABO_3$ systems that exhibit octahedral rotations,[17] where the B-O-B bond *angle* governs the hopping amplitude, in the present work, it is controlled solely via the bond length variation. The bandwidth (W) is proportional to the hopping amplitude; thus, controlling the hopping amplitude is a straightforward and powerful tool to shape the electronic structure of correlated materials.

## 2.2. Symmetry breaking of the V-3*d* t$_{2g}$ states

In addition to bandwidth control, the strain-induced tetragonal distortion increases the correlation strength via crystal-field splitting.[7,34,35] The unstrained SVO *3d$^1$* system has threefold degenerate t$_{2g}$ states at the ground state ($d_{xy}$, $d_{xz}$, $d_{yz}$). Under compressive (tensile) strain, the out-of-plane (in-plane) states, $d_{xz}/d_{yz}$ ($d_{xy}$), are stabilized (**Figure 2a**). Namely, for a fixed integer occupation of n = 1, where n is the number of electrons per site, the number of degenerate bands (N) in the ground state is reduced from N = 3 for an unstrained SVO, to N = 2 or N = 1 for compressive and tensile strained SVO (respectively). This reduction in N, namely the reduction of the available delocalized states, is expected to increase the correlation strength for both strain types.[7]

To establish this strain-induced partial alleviation of the SVO t$_{2g}$ degeneracy, we employ x-ray linear dichroism (XLD) measurements at the V L$_{2,3}$ edge. The samples were measured with two different x-ray linear polarizations: (i) parallel to the film's plane ($I_\parallel$) and (ii) at 30° from the film's normal ($I_\perp$, for simplicity), where most of the signal originates from the perpendicular polarization (**Figure 2b**). The measured signals, $I_\parallel$ and $I_\perp$, are roughly proportional to the corresponding in-plane and out-of-plane unoccupied states. Therefore, XLD, defined as the difference between the two intensities ($I_\perp - I_\parallel$), is sensitive to any preferred occupation in the system. As shown in **Figure 2c**, the XLD signal exhibits a remarkable *mirror* *inversion* between compressive (SVO-LAO) and tensile (SVO-LSAT, SVO-STO) strain states, indicating the expected inverse orbital occupation between biaxial tensile and compressive strain.

The XLD qualitative comparison demonstrates how the application of biaxial strain controls the SVO preferred occupation. In other words, the energy lowering of the $d_{xy}$ (tensile) or the $d_{xz}/d_{yz}$ (compressive) states is governed by the tetragonal distortion. These findings are in line with previous works demonstrating strain-controlled preferred occupations.[36–38] While theoretically predicted for SVO,[39] to our knowledge, this is the first experimental demonstration of strain-controlled preferred occupation of a Mott-Hubbard system, providing direct evidence of the straightforward control over the electronic structure.

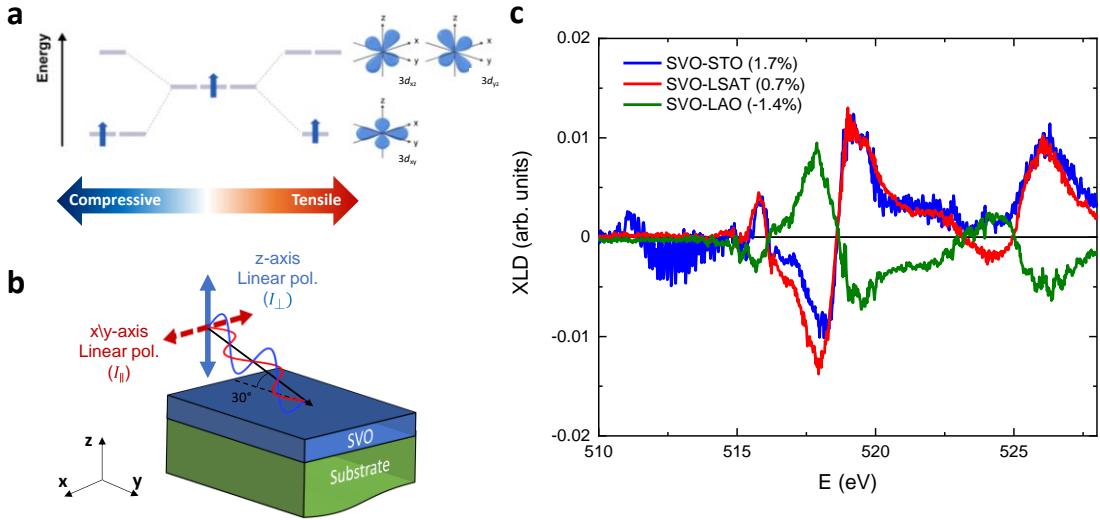

**Figure 2.** Strain control of the preferred occupation. (a) Schematic plots of the degenerate $3d^1$ $t_{2g}$ states under both signs of biaxial strain. (b) Schematic illustration of the x-ray absorption spectroscopy measurement at two linear polarizations. The grazing angle between the x-ray beam to the xy plane is 30°. Thus, $I_\parallel$ probes the in-plane completely, while $I_\perp$ probes mainly the out-of-plane with a small component of the in-plane. (c) Normalized x-ray linear dichroism (XLD) calculated as $I_\perp - I_\parallel$ taken in x-ray excited optical luminescence (XEOL) mode at the V $L_{2,3}$ edge. The x-ray absorption spectroscopy (XAS), defined as $I_\perp + I_\parallel$, and the data analysis are discussed in the Supporting Information and presented in Figures S3 and S4, respectively.

## 2.3. Strain effect on the electronic structure

We now address the question, how does the concurrent modulation of the hopping amplitude and the crystal field splitting govern the correlation strength? For compressive strain, we expect the contributions of the hopping amplitude increase to counteract the crystal field split (N=3→2), altogether exhibiting a negligible contribution to the correlation strength and the electronic structure. Conversely, for tensile strain, we expect both contributions to increase the correlation strength, and drive the system toward its insulating state.

To elucidate how the correlation strength is affected by the epitaxial strain, we study the electronic structure by employing optical conductivity analysis.[40–42] The frequency-dependent optical conductivity is calculated from the complex dielectric function using the relation[43] $\sigma(\omega) = -i\varepsilon_0[\varepsilon(\omega) - 1]\omega$, where ω is the frequency, $\varepsilon_0$ is the vacuum dielectric constant, and $\varepsilon$ is the complex dielectric function (extracted from spectroscopic ellipsometry measurements,[44] Figure S5). To a first approximation, the real part of the optical conductivity (hereafter referred to as the optical conductivity) is proportional to excitations within the single-particle spectral weight of the band structure.[45] Thus, we can use the optical conductivity spectrum to understand the spectral weight redistribution stemming from correlation strength modulation.

The optical conductivity of the strained SVO films is presented in **Figure 3a** (and Figure S6). We assign the features to the known electronic structure of SVO to reveal the role of strain. The Drude peak (peak A) originates from the metallicity of the SVO films, corresponding to intraband transitions of the valence electrons within the conduction band of the V-3$d$ states. The extrapolated conductivity at ω = 0 and the optical mobility are similar to the measured dc conductivity and Hall mobility (Figures S7a and S7b, respectively), further validating the optical analysis. The effective mass of the strained SVO films (Figure S7c), obtained from the Drude peak (see Supporting Information for details), exhibits an increase going from compressive to tensile strain, attesting to the expected increase in correlation strength.

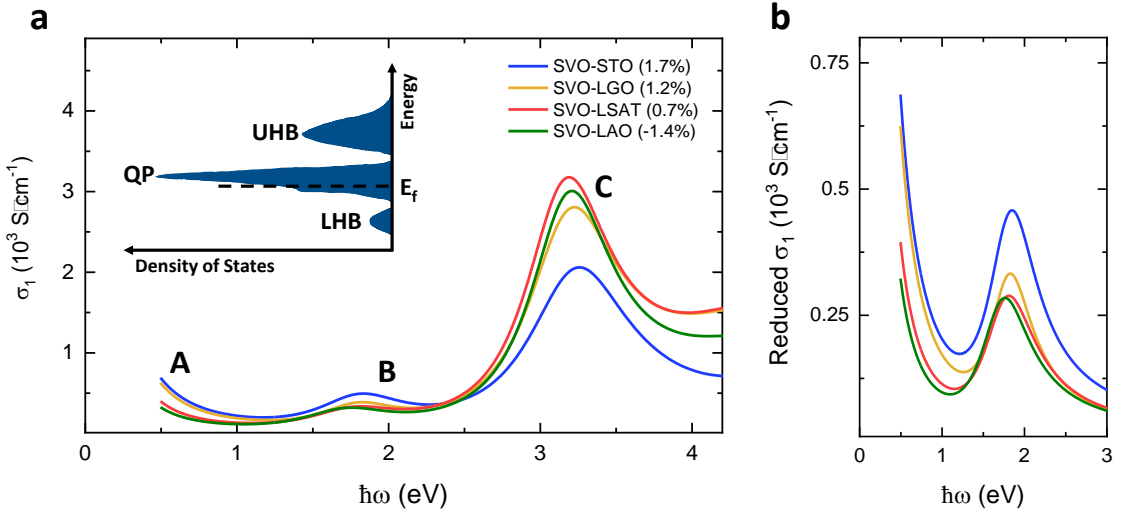

**Figure 3.** (a) Room temperature optical conductivity of SVO under tensile and compressive strain. The metallic spectra exhibit a Drude tail (peak A), a weak feature attributed to the U/2 energy (peak B), and a large feature stemming from the CT excitations (peak C). The inset shows a schematic illustration of the SVO V-3*d* density of states featuring the QP peak and Hubbard bands. (b) Reduced optical conductivity, composed of only excitations within the V-3*d* band (peaks A + B). We subtracted the CT contribution by eliminating their corresponding oscillators for the imaginary part of the dielectric function. Additional information is available in the Supporting Information and Figures S8 and S9.

As a correlated metal, the SVO 3*d* band presents a three-peak structure (Figure 3a insert, following Ref. [24]) featuring additional excitations. As measured and theorized earlier,[17,41,45] the optical conductivity spectra consist of two additional transitions within the V-3*d* band: (i) at about U/2 energy (at ~1.8 eV) corresponding to the transition from the occupied LHB (O-LHB) to unoccupied QP states, and from occupied QP states to the unoccupied UHB (UO-UHB), and (ii) at the U energy (~3.5 eV) corresponding to the transition from the O-LHB to the UO-UHB. Thus, we assign the weak feature at ~1.9 eV (peak B) to the expected U/2 excitation energy, where we ascribe it mainly to excitations from the O-LHB to the UO-QP peak, due to the high spectral weight of the latter. In addition, the SVO O 2*p* states are calculated to be ~3 eV below the Fermi level,[7,24] resulting in a charge-transfer (CT) excitation energy of ~3.5 eV for O 2*p* →V 3*d* transition.[41] Therefore, the feature at ~3.2 eV (peak C) can stem from both U and CT transitions; however, owing to the high O 2*p* spectral weight, the CT feature will overshadow the U peak, making it the prominent component of this feature.

Having explained the optical spectra, we now address the strain-dependent spectral weight evolution. We highlight the distinct change in peaks B and C intensities as the SVO approaches the highest tensile strain (SVO-STO, 1.7%). While the intensity of peak C decreases with increasing strain, the intensity of peak B increases (as illustrated in **Figure 3b**). This behavior can be explained by spectral weight redistribution of the 3*d* band, where the incoherent LHB and UHB spectral weights increase at the expense of the coherent QP peak, as the SVO approaches the Mott insulating state. Hence, tensile strain is shown to increase the SVO correlation strength. We emphasize that this spectral weight redistribution is governed by the magnitude of the biaxial tensile strain, causing a concurrent hopping amplitude decrease and crystal field splitting. Our observation of spectral weight reduction with increasing tensile strain around ~3.2 eV agrees with a previous work, where chemical pressure (Ca substitution) was applied as means to control the bandwidth.[17]

Therefore, utilizing epitaxial strain as the bond length control parameter, the spectral weight can be moved around the Mott landscape in a thin-film scenario, which is important for device implementation. This approach highlights a useful tuning parameter for leveraging Mott physics for new electronic devices, driving them closer toward practical realization.

**2.4. Tetragonal distortion without octahedral rotation**

Some perovskites will accommodate the epitaxial strain via rotation of the BO$_6$ octahedra in addition to changes in the atomic bond distances.[46] If octahedral rotation occurs, the in-plane hopping parameter decreases (increases) under compressive (tensile) strain due to the distortion of the B-O-B bond angle.[16] Conversely, here the strain is expected to induce a modulation of the atomic bond length rather than the angle. In order to rule out the bending of the V-O-V bond angle off its nominal 180°, scanning transmission electron microscopy (STEM) is employed. High-angle annular dark-field (HAADF, **Figures 4a** and **4b**) and annular bright-field (ABF, **Figure 4c**) STEM images along the [100] direction are shown for compressively strained SVO film (SVO-LAO, -1.4%). The results indicate no systematic shift of the V columns along the out-of-plane direction. The electron diffraction pattern of the SVO film is shown in the inset of Figure 3b, where half-integer diffractions are absent, further validating the real-space linear bond observation. A complimentary micrograph for tensile strained SVO film (SVO-LSAT, 0.7%) is presented in Figure S10. Furthermore, using a similar approach, Wang et al. ruled out octahedral rotations for an SVO film on an STO substrate (SVO-STO, 1.7%).[47]

In addition to ruling out octahedral tilts or rotations in the strained SVO films, the STEM analysis further confirms the expected tetragonal distortion in real-space. The strain-induced tetragonal distortion of the unit cell is observed for SVO/LAO via the distortion of the $VO_6$ octahedra. We measured four inner angles of the oxygen octahedra (designated as O1, O2, O3, and O4). It can be seen that the two horizontal angles (O2 and O4) are larger than the two vertical angles (O1 and O3). The average angle of O1 and O3 is $85.4 \pm 2.3°$, whereas the average angle of O2 and O4 is $94.5 \pm 1.9°$. These inner octahedron angles are comparable to those calculated using the tetragonal lattice parameters (Table S1), which are $88.8°$ for O1 and O3, and $91.2°$ for O2 and O4.

We note an additional possible strain accommodation mechanism, by the formation of defects such as vacancies and dislocations. However, the high structural quality and RRR values (see Supporting Information for details) suggest that the small possible concentration of defects carries a negligible effect onto the SVO electronic properties.

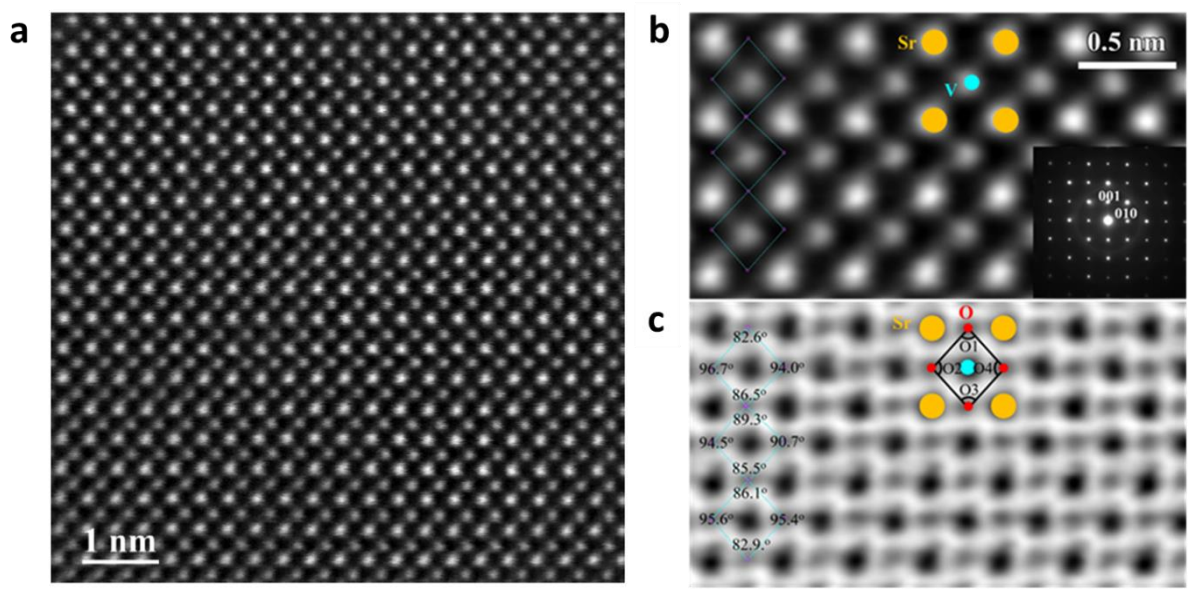

**Figure 4.** Cross-section high-resolution STEM micrographs of a compressively strained SVO film on an LAO substrate, imaged with HAADF (a) and (b), and ABF (c) along the [100] zone

axis. The electron diffraction pattern is shown in the inset of (b), showing the absence of half-integer diffractions. Oxygen octahedral angles are evaluated from the ABF STEM image (see Experimental for details).

## 3. Conclusions

SVO provides a simple testbed and model for electron correlation. We harnessed this simplicity to derive a clear picture of the effect of biaxial strain on the correlation strength and the resulting electronic structure. Using a thorough structural analysis, we show that in this nominally-cubic system, the epitaxial strain directly translates to V-O bond length modulation via tetragonal distortion without changing the V-O-V bond angle. In addition to the variation of the hopping amplitude, the bond length modulation also results in crystal field splitting of the nominally-degenerate V-3$d$ $t_{2g}$ orbitals; both contributions are considered to concurrently affect the correlation strength and the V-3$d$ band structure. We therefore demonstrate strain-dependent preferred orbital occupancy in a Mott-Hubbard system by exploiting the orbital degeneracy reduction. The change in correlation strength is manifested in the SVO V-3$d$ band spectral weight redistribution. Tensile strain is shown to further drive the system towards an insulating state, a consequence of the increased correlation strength. We thus demonstrate a straightforward approach for modulating the electronic structure of this correlated model system. Applying these methods to more complex systems, particularly near an electronic phase transition, is expected to yield robust effects and large changes in the electronic structure, providing an efficient tuning parameter for harnessing correlated materials and Mott physics towards novel practical solid state electronic devices and circuits.

## 4. Experimental Section

*Sample Preparation*: SVO films (60 u.c.) were epitaxially grown on (001)-oriented (cubic\pseudocubic) LAO, LSAT, STO, and LGO substrates (CrysTec GmbH). This thickness was chosen to reduce the contribution of possible surface effects[48] while ensuring the coherent epitaxy of the films. The growth was done using oxide molecular beam epitaxy (MBE) at a background pressure of molecular oxygen of $\sim 5 \times 10^{-7}$ Torr and substrate temperature of 1000°C.[30] Substrate preparation is described in the Supporting Information. After growth, the samples were cooled to room temperature under the same oxygen ambient as the growth.

*Structural Characterization*: 2θ/ω scans and RSMs were performed with x-ray diffraction (XRD) using a Rigaku SmartLab with a 2-bounce incident monochromator.

*Scanning Transmission Electron Microscopy*: Cross-sectional TEM sample was prepared by a focused ion beam (FIB) with Ga+ ions. The final milling was done with 5 keV Ga+ ions to reduce FIB-induced damage. A cross-sectional TEM sample with its plane normal tilted ~ 22.5º from the [100] and [110] directions was prepared, which is within the tilting range of the TEM sample holder. By tilting +/- 22.5º, we obtained a set of atomic resolution images and electron diffraction patterns along both the [100] and [110] directions from the same area of the sample to examine octahedra tilts or rotations. The JEOL ARM 200CF at the Brookhaven National Laboratory equipped with double aberration correctors was used to obtain HAADF and ABF STEM images. The ranges of the collection angles for HAADF and ABF are 67 ~ 275 and 11 ~ 23 mrad, respectively. The convergence angle for the imaging electron beam was 21.2 mrad. Oxygen octahedra angles are measured in the ABF STEM image after refining the oxygen positions using the Oxygen Octahedra Picker.[49]

*X-ray Absorption Spectroscopy*: The XAS spectra were acquired at the Swiss Light Source (SLS) on the XTreme beamline.[50] The measurements were performed using XEOL detection mode[51] under zero magnetic field in linear horizontal and vertical polarizations. The SVO-LAO and SVO-LSAT samples were measured at room temperature, while the SVO-STO sample was measured at 120 °K to increase luminescence from the STO substrate.

*Spectroscopic Ellipsometry (NIR+VIS+UV)*: Ellipsometry data was acquired using a dual rotating compensator ellipsometer (RC2 ellipsometer, J. A. Woollam Co.). Data were acquired at 1058 wavelengths from 193 to 2500 nm (0.496 to 6.425 eV), at incident angles of 55, 60, 65, 70, and 75°. Focusing attachments reduced the beam diameter to approximately 300 µm.

**Supporting Information**

Supporting Information is available from the Wiley Online Library or from the author.

**Acknowledgements**

This work was funded by the Israeli Science Foundation (ISF Grant No. 1351/21). Sample processing was done at the Technion's Micro-Nano Fabrication & Printing Unit (MNF&PU). Sample characterization was done with partial support from The Russell Berrie Nanotechnology Institute (RBNI) and The Nancy and Stephen Grand Technion Energy Program (GTEP). The work at the Brookhaven National Laboratory was supported by the Materials Science and Engineering Division, Office of Basic Energy Sciences, of the U.S. Department of Energy under Contract No. DESC001274. FIB sample preparation was performed at the Center for Functional Nanomaterials, Brookhaven National Laboratory. A.Z. acknowledges the financial support by the Swiss National Science Foundation (SNSF) under Project No. 200021_169467. We thank Dr. Vladimir Kalnisky and Mr. Brajagopal Das for fruitful discussions and their critical reading of the manuscript.

Supporting Information

**Bandwidth Control and Symmetry Breaking in a Mott-Hubbard Correlated Metal**

*Lishai Shoham, Maria Baskin, Tom Tiwald, Guy Ankonina, Myung-Geun Han, Anna Zakharova, Shaked Caspi, Shay Joseph, Yimei Zhu, Isao H. Inoue, Cinthia Piamonteze, Marcelo J. Rozenberg, and Lior Kornblum*

1. **Sample Preparation**

All the substrates used here are 5×5×0.5 mm$^3$ supplied by CrysTec GmbH. SrTiO$_3$ (STO) substrates were TiO$_2$ terminated using the HF method.[1] LaAlO$_3$ (LAO) substrates were annealed in air at 1000°C for 2.5 hours. (LaAlO3)$_{0.3}$(Sr$_2$TaAlO$_6$)$_{0.7}$ (LSAT) substrates were annealed in air at 1300°C for 2.5 hours. SVO films for XAS and STEM analysis were grown on as-received LAO and LSAT substrates. LaGaO$_3$ (LGO) substrates were used as received.

2. **Structural and Electronic Characterization**

**2.1. X-ray Diffraction (XRD).** 2θ/ω scans around the (002) substrate peak were performed for all strained SVO films (Figure S1).

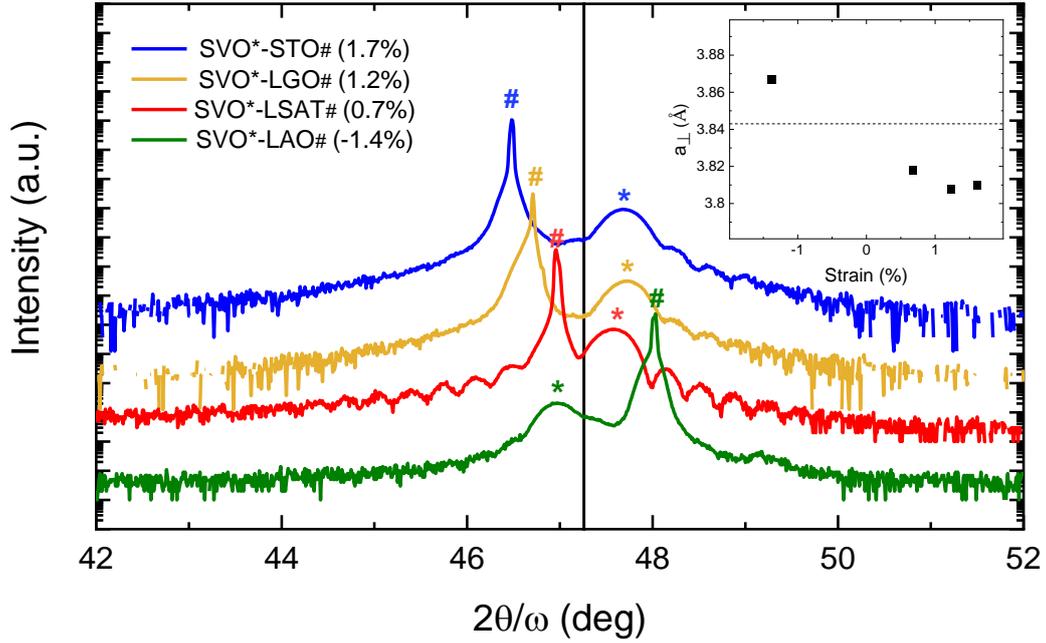

**Figure S1.** High resolution 2θ/ω x-ray diffraction scans around the (002) Bragg peak of each substrate. The shift of the SVO out-of-plane lattice parameter is illustrated in the inset, featuring the out-of-plane lattice parameter as a function of the biaxial strain.

The XRD data were fitted using GlobalFit by Rigaku, considering film thickness, film out-of-plane lattice parameter ($a_{\perp,SVO}$), and substrate lattice parameter ($a_{sub}$). The fit parameters are summarized in Table S1. The tetragonal distortion manifests in the elongation (shortening) of the out-of-plane lattice parameter with the increase of the compressive (tensile) strain. We note that the strained SVO film on STO substrate exhibits a deviation in this trend, as shown in the inset of Figure S1. This small deviation can be explained by the presence of a small concentration of point defects formed to accommodate the high tensile strain. Nevertheless, the transport measurement in the following section confirms that this small concentration of defects did not degrade the high quality of the film. We emphasize that the films exhibit high RRR values, meaning that the small increase in the point defect density is not expected to produce meaningful changes in the band structure.

Using the in-plane lattice parameter of the substrate (for these coherent films) and the extracted SVO out-of-plane lattice parameter, one can calculate the four inner angles of the $VO_6$ octahedra (designated as O1, O2, O3, and O4, see main text). The calculated angle variation indicates a ±1.5° deviation from the unstrained 90° inner angle for highly tensile and compressive strained SVO films.

**Table S1.** Summary of the parameters used for the XRD fitting results, and the calculated inner angles of the $VO_6$ oxygen octahedra.

|  | $a_{sub}$ (Å) | $a_{\perp,SVO}$ (Å) | Thickness (nm) | $\Theta_{1,3}$ (deg) | $\Theta_{2,4}$ (deg) |
|---|---|---|---|---|---|
| **SVO-STO** | 3.905 | 3.810 | 27 | 88.6 | 91.4 |
| **SVO-LGO** | 3.890 | 3.808 | 27 | 88.8 | 91.2 |
| **SVO-LSAT** | 3.868 | 3.818 | 26 | 89.3 | 90.7 |
| **SVO-LAO** | 3.790 | 3.867 | 29 | 91.2 | 88.8 |

**2.2. Transport Measurements.** Sheet resistance and Hall resistance were measured in the Van der Pauw geometry using a Quantum Design Physical Properties Measurement System (PPMS). The contact to the films was made using Al wire, which was directly wedged to the bare surface of the film. Sheet resistance was measured in the temperature range of 2 - 300 °K. Figure S2 illustrates the resistivity as a function of the temperature for the different strained SVO films.

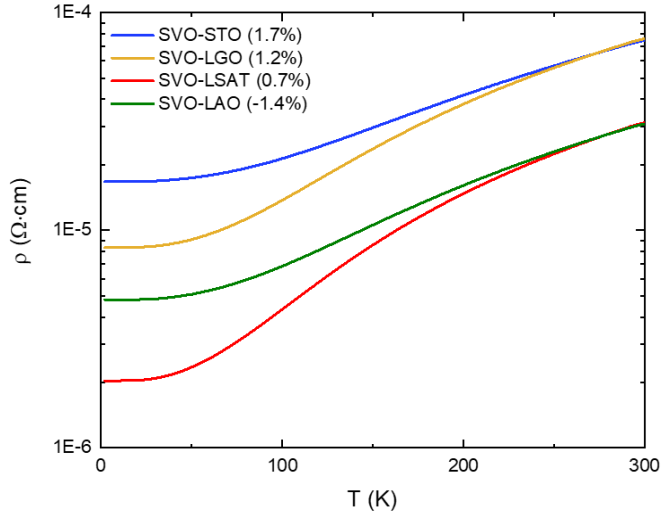

**Figure S2.** Temperature-dependent resistivity for the strained SVO films.

Table S2 summarizes the residual resistivity ratio (RRR), defined as $\rho_{300K}/\rho_{2K}$, and the room temperature conductivity for the SVO films. Hall resistance measured at room temperature between ± 4 T was used to extract the electron density and mobility. The Hall resistance was linear with magnetic field variation, presenting a negative slope indicating electrons as the carrier density.

**Table S2.** Summary of the transport properties as extracted from the transport measurements.

|  | RRR | σ @ 300K  ($10^4$ S·cm$^{-1}$) | n @ 300K  ($10^{22}$ cm$^{-3}$) | µ @ 300K  (cm$^2$·S$^{-1}$V$^{-1}$) |
|---|---|---|---|---|
| **SVO-STO** (L303) | 4.5 | 1.34 | 1.9 | 4.3 |
| **SVO-LGO** (L298) | 9.1 | 1.32 | 2.3 | 3.5 |
| **SVO-LSAT** (L263) | 15.3 | 3.20 | 2.1 | 9.6 |
| **SVO-LAO** (L59) | 6.4 | 3.24 | 1.9 | 10.9 |

## 3. X-ray Absorption Spectroscopy (XAS).

**3.1. XAS measurements** were measured in x-ray excited optical luminescence (XEOL) mode[2] at the V $L_{2,3}$ edge for the different strained films. The x-ray was in grazing incidence, with the sample surface at 30º with the incoming x-ray beam (Figure 2b). The measurements were conducted under zero magnetic field in linear vertical and 30° to the linear horizontal polarizations with respect to the film's normal (LV and LH, respectively). The SVO-LAO and SVO-LSAT samples were measured at room temperature, while the SVO-STO sample was measured at 120 °K to increase luminescence from the STO substrate. However, the low-temperature measurement is not expected to affect the results, as seen by the close agreement of the 120 °K to the 300 °K SVO-LSAT data (Figure 2c), which is also tensely strained. The XAS and x-ray linear dichroism (XLD) for the different strained film spectra are presented in Figure S3. We do not address the adjacent O K-edge due to the strong contribution from substrate oxygen.

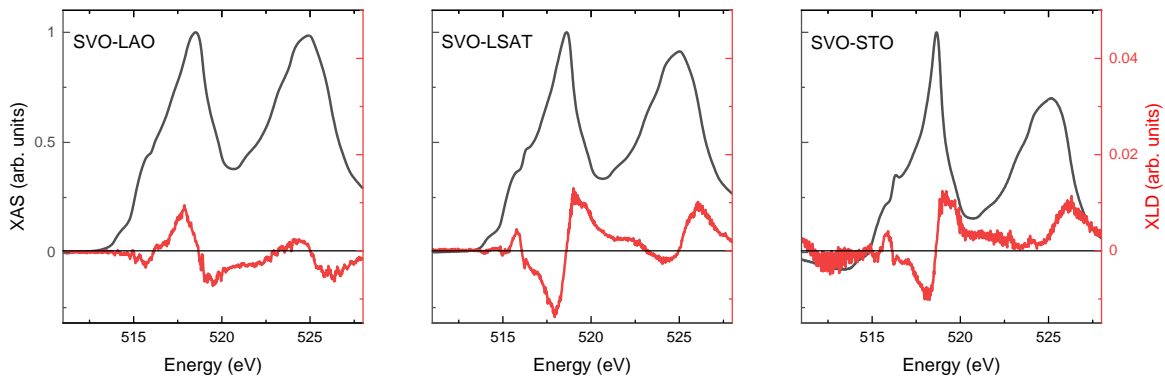

**Figure S3.** Normalized XAS (black) and XLD (red) at the V $L_{2,3}$ edge spectra for the strained SVO films grown on LAO, LSAT, and STO substrates measured in XEOL mode. The STO measurements were done at 120° K to increase substrate luminescence. The XAS is calculated as LH+LV, and the XLD is calculated as LH–LV.

**3.2. Obtaining the absorption results from the XEOL measurement.** First, we measured the sample at LH and LV polarizations at least four times each. After eliminating outliers, we separately averaged the LH and LV results and normalized them at the pre-edge to 1, as shown in Figure S4a. Then, we used the Beer-Lambert law for linear absorption to achieve the LH and LV absorption signal. The results are presented in Figure S4b.

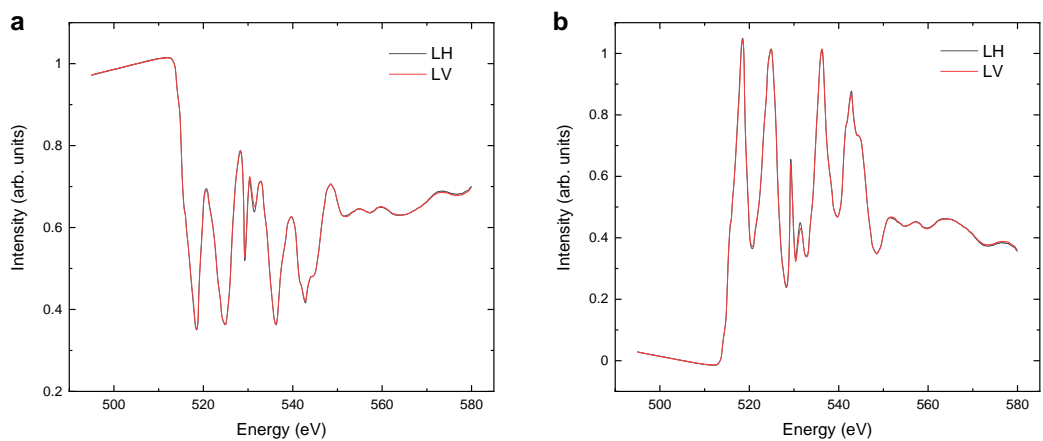

**Figure S4.** An example of obtaining the absorption results from the XEOL measurement of the SVO-LAO sample. (a) Averaged XEOL intensity as recorded for LH (black) and LV (red). The data were normalized to 1 at the pre-edge. (b) The calculated absorption for LH (black) and LV (red) using Beer-Lambert law for linear absorption.

**3.3. Normalizing the XAS and XLD results:** we calculated the XAS to be LH+LV and the XLD to be LH-LV. For the XAS, we subtracted a linear background fitted to the pre-edge and normalized it to 1 at $L_3$ maximum. The XLD was normalized by the same factor used to normalize the XAS (no background was subtracted from the XLD).

## 4. Spectroscopic Ellipsometry (SE).

**4.1. SE measurements.** All strained SVO films and bare substrates were measured using *ex-situ* SE at room temperature. The ellipsometry spectra (in ψ, Δ) were collected using focusing probes to reduce the beam diameter to approximately 300 µm, which allowed the illuminated area to easily fit on the relatively small samples. No other sample preparation was necessary.

**4.2. SE modeling**. The complex dielectric function ($\varepsilon = \varepsilon_1 + i\varepsilon_2$), presented in Figure S5, was obtained by fitting the ellipsometry spectra collected over the near IR to UV range. The spectra were fitted to a model consisting of a semi-infinite substrate / SVO film using CompleteEASE® software (J. A. Woollam Co.). For some of the samples, an additional surface roughness layer was added. The goodness of the fit is quantified by using the Root Mean Squared Error[3].

The substrates' dielectric functions were determined from separate measurements of the bare substrates. The LGO substrate is anisotropic; therefore, it was modeled with a uniaxial dielectric tensor. During measurement, this sample was oriented such that the optic axis was parallel to the plane of incidence, meaning there is no significant p-to-s conversion, and the standard ellipsometry measurement provides accurate results.

First, the film layer was fitted using a Kramers-Kronig consistent B-spline function and a fixed thickness extracted from x-ray reflectivity (XRR) measurements (when the surface roughness layer was added, the total thickness was fixed). Then, the parameterization of the layer $\varepsilon_2$ was done using the following five oscillators:[4]

  I. Drude peak with fixed carrier density (taken from Table S2).
  II. Lorentz function at the U/2 energy (~1.8 eV).
  III. Tauc-Lorentz function at the CT($t_{2g}$) energy (~3.2 eV).
  IV. Lorentz function at the CT($e_g$) energy (~5.5 eV).
  V. Additional Lorentz function at the energy ~4.5 eV.

The contribution from the U excitation was neglected, as elaborated in the main text.

Each of these five complex functions is Kramers-Kronig consistent, meaning that the $\varepsilon_1$ of the layer has the correct line shape within a constant offset, which is determined by fitting a constant $\varepsilon_\infty$. For SVO-LSAT and SVO-STO, an additional surface roughness layer was added.

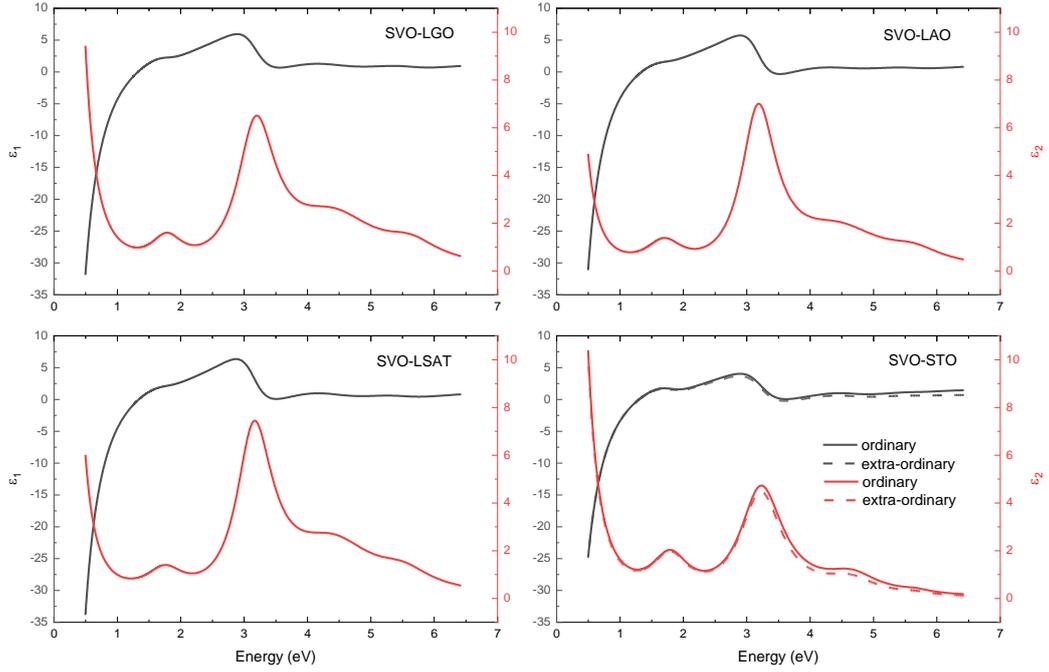

**Figure S5.** Real and imaginary dielectric functions spectra (black and red, respectively) of strained SVO films grown on LGO, LAO, LSAT, and STO substrates. The spectra were extracted from room-temperature spectroscopic ellipsometry measurements. For SVO-STO we used a uniaxial model, as elaborated below.

We found that using a uniaxial model ($n_x, n_y \neq n_z$) significantly improves the fit for the SVO-STO sample. In this model, the in-plane dielectric function (ordinary) is determined as elaborated above, whereas the in-plane dielectric function models the out-of-plane dielectric function (extra-ordinary) by adding a small difference value to the index $n(\lambda)$. The SVO-STO optical conductivity in Figure 3a is therefore based on the in-plane component. We note that the epitaxial strain is expected to create an optical anisotropy between the in-plane and out-of-plane dielectric functions. However, the sensitivity to this change is negligible due to the low thickness of the films (compared to the measurement wavelength). As it appears, there is a small sensitivity under high tensile strain (SVO-STO). While the anisotropic model improves the goodness of the fit, the fit result and trend are unchanged whether an isotropic or an anisotropic model is used.

We further considered the effect of the near-surface region (NSR) observed on SVO films[5,6] on the reliability of the model. To that end, we modeled the SVO layer as elaborated before (4.2 SE modeling) and added an independent B-spline surface layer with a fixed thickness of 4 nm.[5] The total thickness (layer + NSR) was fixed to the thickness measured via XRR. As a result, the additional surface layer did not improve the fit substantially, and the layers' optical constants were similar to the previous models. Therefore, we conclude that there is minor sensitivity to the presence of the surface region.

**4.3. The optical conductivity** was calculated using the relation:

$$\sigma(\omega) = -i\varepsilon_0[\varepsilon(\omega) - 1]\omega \qquad (1)$$

where $\varepsilon_0$ is the vacuum dielectric constant, $\varepsilon$ is the complex dielectric function (see Figure S5), and $\omega$ is the frequency. The energy-dependent optical conductivity, as calculated for the full measured spectrum range, is presented in Figure S6. For metals, optical conductivity consists of intraband transitions of free electrons within the conduction band (Drude), and interband transitions of bound electrons from lower energy bands to the unoccupied part of the conduction band. An example of the latter is the transition from the occupied O $2p$ band to the unoccupied TM-$3d$ band, named charge transfer (CT).

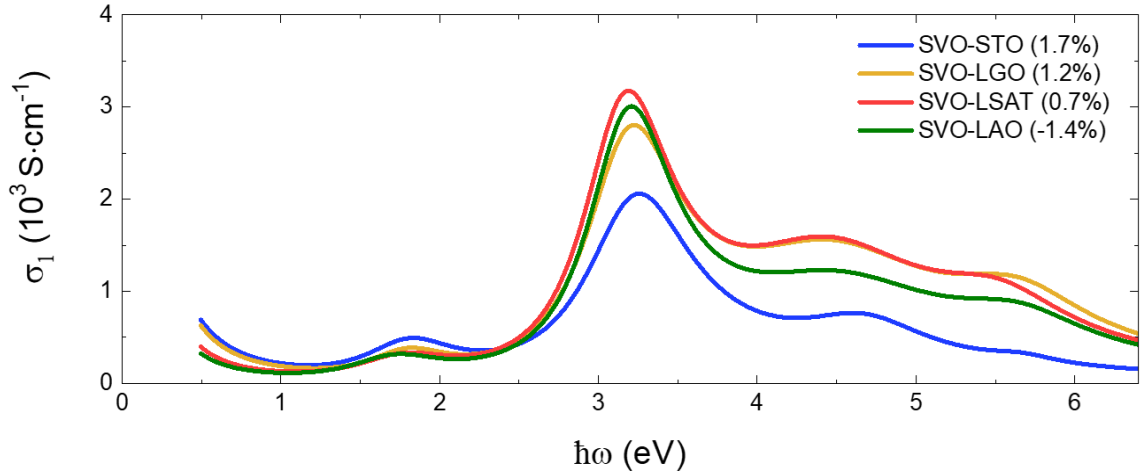

**Figure S6.** Optical conductivity of SVO under tensile and compressive strain. Peaks at energies above 4.2 eV are ascribed to additional excitations from the broad O $2p$ states to the unoccupied V states (e.g., V-$3d$ $e_g$ states).[4]

The reduced optical conductivity (Figure 3b) was calculated from the summation of oscillators I and II (4.2 SE modeling), representing the Drude peak and U/2 excitation. Thus, the reduced optical conductivity is composed only of the V-3$d$ intraband excitations.

**4.4. The Drude model.** From the modeling of the Drude peak using the CompleteEASE software, we can extract the optical mobility (Figure S7a), optical dc conductivity (Figure S7b), and effective mass (Figure S7c). The similar values for µ and $\sigma_{dc}$ extracted from the optical and transport measurements attest to the reliability of the optical analysis. The effective mass increase going from compressive to tensile agrees with our theoretical expectations that the increase in V-O bond length increases the correlation strength.

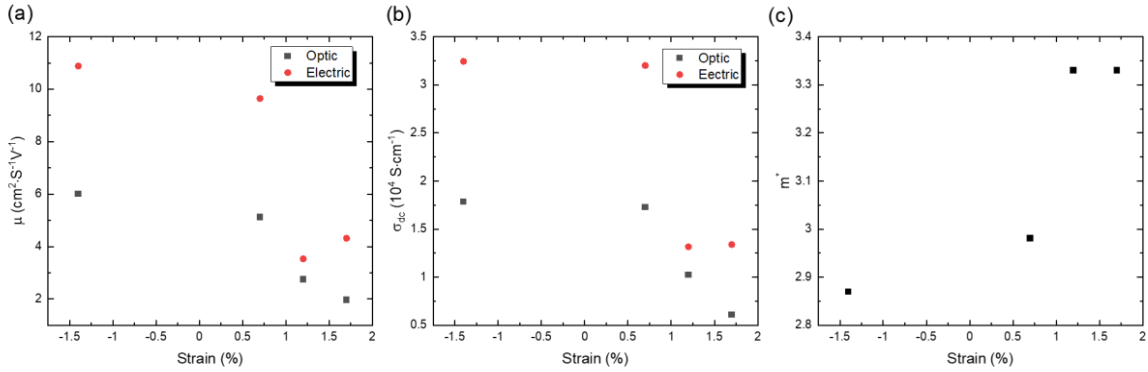

**Figure S7.** SVO mobility (a) and dc conductivity (b) as extrapolated from the optical measurements (black) or electrical measurements (red), respectively. (c) The SVO effective mass was extrapolated from the optical measurements. The data were acquired at room temperature.

Using the optical values, the Drude peak can be extrapolated via the relation (Figure S8):[7]

$$\varepsilon_{Drude}(\omega) = \frac{-q^2 N \mu}{\varepsilon_0(\mu m^* m_e \omega^2 + iq\omega)} \quad (2)$$

where $\omega$ and $\varepsilon_0$ are the frequency and vacuum dielectric constants, respectively. N is the carrier density, and q, $m_e$, $m^*$, and µ are the electron charge, rest mass, effective mass, and mobility, respectively. Extrapolation of the measurement to $\omega \to 0$ is done using the measured carrier densities (Eq. 2, Table S2).

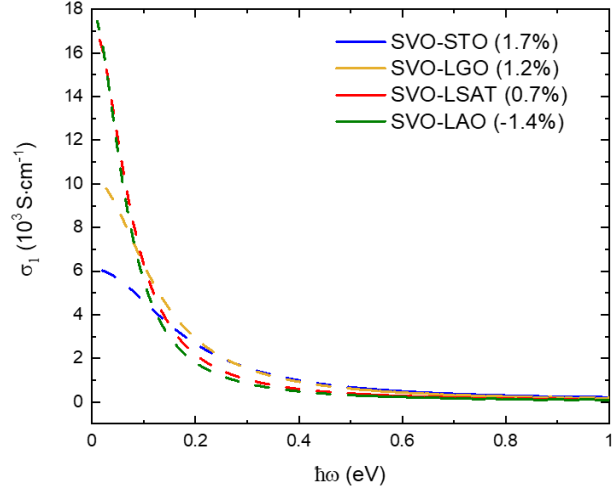

**Figure S8**. The Drude peak optical conductivity in the measured range (solid) and as extrapolated using equation 2 (dashed).

**4.5. Effective number of electrons.** By considering the sum rule of the conductivity spectra, we can estimate the effective number of free carriers ($N_{eff}$) defined as:

$$N_{eff} = \frac{2m}{\pi e^2 V} \int_0^\omega \sigma(\omega')d\omega' \qquad (3)$$

Here, m is taken as $m_0$, the free electron mass, V is the number of vanadium atoms per unit volume, and ω is the frequency. The effective number of free carriers is proportional to the number of electrons involved in the optical excitations up to frequency ω. $N_{eff}(\omega)$ for the strained SVO films is presented in Figure S9.

Theoretically, one free electron is expected per SVO unit cell. However, we have assumed that the electron mass equals the free carrier mass (m = $m_0$), whereas it is expected to be higher (m > $m_0$) due to the electron correlation. When we consider the extrapolated effective mass (~3.0, as extrapolated from the SE. See Figure S7c), $N_{eff}$ equals one at about ~1 eV. This result agrees with our understanding that the majority of the V 3*d* single-particle spectral weight is within the quasiparticle peak.

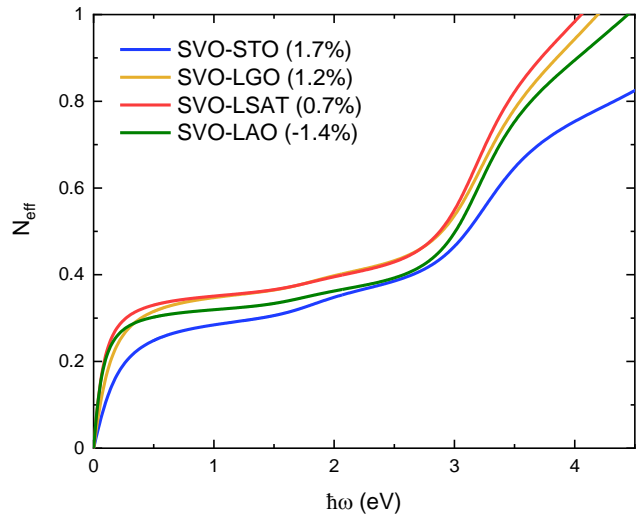

**Figure S9.** Effective electron number per vanadium atom for the different strained SVO films.

## 5. Scanning Transmission Electron Microscopy.

### 5.1. Tensile Strain: SVO-LSAT Sample.

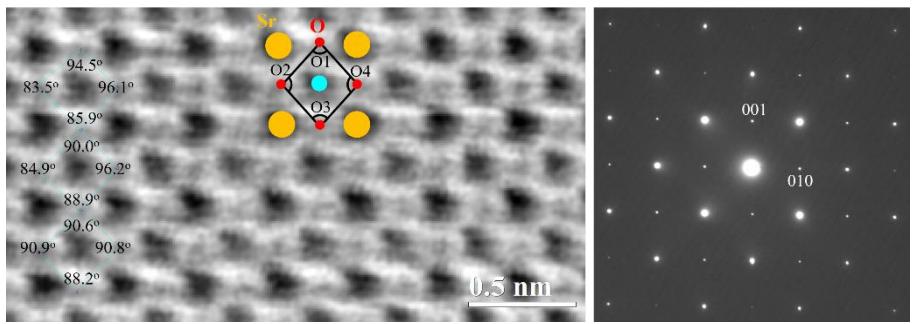

**Figure S10.** Cross-section high-resolution scanning transmission electron microscopy (STEM) micrographs of a tensile strained SVO film on LSAT substrate imaged with annular bright-field (ABF) along the [100] axis. It shows the oxygen octahedra inner angle measurements and selected area electron diffraction patterns from the SVO and LSAT substrate. The measured angles are shown on the selected octahedra. The measurements show that the average horizontal angles (O2 and O4) are $90.4 \pm 4.9°$ while the vertical angles (O1 and O3) are $93.0 \pm 2.6°$.